\documentclass[prl,aps,preprint,showpacs]{revtex4}

\usepackage[dvips]{graphicx}
\usepackage{amsmath}

\begin{document}

\title{$1/f$ noise in chemical reactions}

\author{A. Fuli\'nski}
\email{fulinski@if.uj.edu.pl}

\author{P.  F.  G\'ora}
\affiliation{M.~Smoluchowski Institute of Physics, Jagellonian
University, Reymonta~4, 30--059~Cracow, Poland}

\date{\today}

\begin{abstract}
A chemical system consisting of two species, one of which evolves
deterministically and independently of the other, which in turn 
is driven by the dynamics of the former and by an additional
multiplicative Gaussian white noise, displays a $1/f$ noise
for intermediate to large frequencies. A~novel mechanism
responsible for the $1/f$ noise is suggested.
\end{abstract}


\maketitle

Although systems exhibiting $1/f^\alpha$ noise, and related systems which display 
power--law type distributions, are abundant in nature, there is
no generally accepted explanation for their origin and ubiquity. Self--organized
criticality \cite{SOC} has been proposed as such an explanation, but it has been
pointed out that the notion of self--organized criticality and $1/f^\alpha$ are
mutually exclusive in many cases \cite{critique}. Recently the emergence
of power--law distributions and long--time temporal correlationshas have been 
explained for example in terms of a generalized logistic systems \cite{Biham}.
More importantly, Davidsen and Schuster in Ref.~\cite{Schuster} have proposed 
a mechanism explaining the origin of $1/f^\alpha$, $\alpha\simeq1$, noise
in the low--frequency regime, which corresponds to very long lasting temporal 
correlations. The systems considered in Ref.~\cite{Schuster} were driven by
a Gaussian white noise (GWN) --- in other words, no preexisting power--law 
distributions or fractal noises have been assumed, even implicitly. Power--law 
distributions have also been shown to result, at the level of the mathematical 
formalism used, from nonlinear transformation between various stochastic variables
for a very wide class of underlying ``fundamental'' distributions~\cite{Vlad}.

In the present Letter we consider a simple chemical system consisting of two 
species. Dynamics of one of the species, A, is subjected to a multiplicative
GWN and is further affected by the dynamics of the other
species, B. The dynamics of B is purely deterministic and is not influenced back
by the dynamics of A. Such a system can be realized experimentally. We show that 
our model system exhibits $1/f$ noise for intemediate to large frequencies.

Consider two chemical species, A and B. Let B decay autocatalytically
and catalyse a decay of A:

\begin{eqnarray}
\label{A}
\mathrm{A} + \mathrm{B} &\mathop{\longrightarrow}\limits_K& 
\mathrm{A}^*\!\downarrow + \mathrm{B}\,,
\\
\label{B} 
\mathrm{B} + \mathrm{B} &\mathop{\longrightarrow}\limits_{k_b}& 
\mathrm{B}^*\!\downarrow + \mathrm{B}\,.
\end{eqnarray}

\noindent The downarrows mean that species $\mathrm{A}^*$, 
$\mathrm{B}^*$ are stable and do not enter any reactions of interest.
If the amounts of reagents are periodically incremented from 
outside

\begin{equation}\label{fluxes}
j_a(t) = a_0\sum\limits_n\delta(t-nT)\,,\quad
j_b(t) = b_0\sum\limits_n\delta(t-nT)\,,
\end{equation}

\noindent the kinetics of reactions (\ref{A})--(\ref{B}) can be expressed as

\begin{equation}\label{kinetyka}
\dot a = -Kab + j_a(t)\,,\quad
\dot b = -k_bb^2 + j_b(t)\,,
\end{equation}

\noindent
where $a$, $b$ are concentrations of A and B, respectively. We further 
assume that the rate at which A decays fluctuates around a deterministic 
value

\begin{equation}\label{noise}
K = k_a + \kappa\eta(t)\,,
\end{equation}

\noindent where $\eta(t)$ is a GWN with 
$\left\langle\eta(t)\right\rangle=0$, 
$\left\langle\eta(t)\eta(t^\prime)\right\rangle=\sigma^2\delta(t-t^\prime)$, 
and all higher correlations factorize. Reactions (\ref{A})--(\ref{B}) with 
the above constrains can be realized experimentally in a flow reactor,
for instance by a photoactivated chemical reaction with a radiation field
acting as a source of noise \cite{Fonseca}; see also \cite{AF2001}
and references quoted therein.

Because of the $\delta$--terms in the fluxes, we solve (\ref{kinetyka})
in a stroboscopic representation. Between the pulses

\begin{equation}\label{b}
b(t) = \frac{b_n}{1+k_bb_n(t-nT)}\,,\quad nT<t<(n+1)T\,,
\end{equation}

\noindent where $b_n= b(t=nT^+)$. The pulses simply increment the 
concentrations by $a_0$ and $b_0$, respectively. Thus 

\begin{subequations}\label{n+1}
\begin{eqnarray}
b_{n+1} &=& b_0 + \frac{b_n}{1+ b_nk_bT}\,,\\
a(t) &=& \frac{a_n}{\left(1+b_nk_b(t{-}nT)\right)^\mu}
\exp\left(-\int\limits_{\smash{nT^+}}^t\!\!\!%
\kappa\eta(t^\prime)b(t^\prime)dt^\prime\!\right),
\nonumber\\&&
nT<t<(n+1)T\,,\\
a_{n+1} &=& a_0 + \frac{a_n}{(1+b_nk_bT)^\mu}E_n\,,
\end{eqnarray}
\end{subequations}

\noindent
where $a_n=a(t=nT^+)$, $\mu=k_a/k_b$ is a ratio of deterministic reaction 
rates and

\begin{equation}\label{En} 
E_n = \exp\left(-\int\limits_{\smash{nT^+}}^{\smash{(n+1)T^-}}\!\!\!%
\kappa\eta(t^\prime)b(t^\prime)dt^\prime
\right)\,.
\end{equation}

The $\{b_n\}$ form a deterministic series convergent to

\begin{equation}\label{binfty}
b_\infty= \frac{b_0}{2}\left(1 +
\sqrt{1+\frac{4}{b_0k_bT}}\right)\,.
\end{equation}

\noindent Note that $b_\infty>b_0$. 
Contrariwise, the series $\{a_n\}$ is stochastic and its realizations need 
not to be convergent; only the series of expectation values 
$\{\left\langle a_n\right\rangle\}$ can. Both $a_n$ and $E_n$ are
random numbers, but since $a_n$ depends only on times prior to $nT$,
$a_n$ and $E_n$ are defined on disjoint intervals and for $\eta(t)$
beging a~GWN, the expectation value of their product factorizes.
We have

\begin{equation}\label{Enaverage}
\left\langle E_n\right\rangle 
=
\exp\left({\textstyle\frac{1}{2}}\kappa^2\sigma^2
\frac{b_n^2T}{1+b_nk_bT}\right)
\end{equation}

\noindent (cf.\ \cite{JSP}), and

\begin{equation}\label{an+1average}
\left\langle a_{n+1}\right\rangle = a_0 + 
\frac{\left\langle a_n\right\rangle}{(1+b_nk_bT)^\mu}
\exp\left({\textstyle\frac{1}{2}}\kappa^2\sigma^2
\frac{b_n^2T}{1+b_nk_bT}\right)\,.
\end{equation}

\noindent In the limit $n\to\infty$ we replace $b_n$ by $b_\infty$ and 
(\ref{an+1average}) yields

\begin{equation}\label{ainfty}
\left\langle a_\infty\right\rangle =
a_0\left[1 - (1+b_\infty k_bT)^{-\mu}
\exp\left({\textstyle\frac{1}{2}}\kappa^2\sigma^2\frac{b_0}{k_b}\right)
\right]^{-1}\,.
\end{equation}

\noindent There is a maximal addmissible noise level

\begin{equation}\label{kappamax}
\kappa^2\sigma^2_{\mathrm{max}} = \frac{2k_a}{b_0}\ln(1 + b_\infty k_bT)
\end{equation}

\noindent beyond which the process diverges. 

We also need to specify a variance of the process $a_n$:

\begin{eqnarray}\label{variance}
v_{n+1}^2 &=& \left\langle a_{n+1}^2\right\rangle - 
\left\langle a_{n+1}\right\rangle^2
\nonumber\\ 
&=&
\frac{v_n^2\left\langle E_n^2\right\rangle + 
\left\langle a_n\right\rangle^2\left(
\left\langle E_n^2\right\rangle-\left\langle E_n\right\rangle^2
\right)}{(1+b_nk_bT)^{2\mu}}\,.
\end{eqnarray}

\noindent It is easy to verify that the series $\{v_n^2\}$ converges for 
$\kappa^2\sigma^2 < \frac{1}{2}\kappa^2\sigma^2_{\mathrm{max}}$
and diverges for higher noise levels. 

Finally we need to examine the autocorrelation structure of the
process $a(t)$. We define

\begin{equation}\label{korelacjadef}
C(t,\tau) = \left\langle a(t)a(t+\tau)\right\rangle -
\left\langle a(t)\right\rangle\left\langle a(t+\tau)\right\rangle\,.
\end{equation}

\noindent For a GWN we obtain

\begin{equation}\label{korelacja1}
C(nT,mT) = 
\left(\left\langle a_n^2\right\rangle - \left\langle a_n\right\rangle^2\right)
\;\prod\limits_{i=0}^{m-1}B_{n+i} \left\langle E_{n+i}\right\rangle\,,
\end{equation}

\noindent where $B_l = (1 + b_lk_bT)^{-\mu}$. Note that all $E_l$'s in
(\ref{korelacja1}) are defined on mutually disjoint intervals and depend on
times later than $nT$. For $n\gg1$, when the process 
reaches its stationary state,  (\ref{korelacja1}) simplifies to

\begin{equation}\label{asymptotycznie}
C(nT,mT) \simeq C_m = v_\infty^2 \left(
\frac{\exp\left(b_0\kappa^2\sigma^2/(2k_b)\right)}{(1+b_\infty k_bT)^\mu}
\right)^m
\end{equation}

\noindent provided that 
$\kappa^2\sigma^2 < \frac{1}{2}\kappa^2\sigma^2_{\mathrm{max}}$.

\begin{figure}
\includegraphics[scale=0.72]{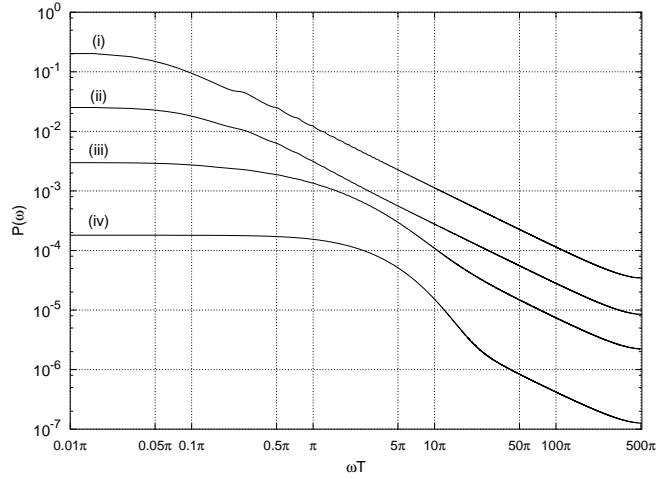}
\caption{Power spectra corresponding to the correlation function (\ref{korelacja3}):
(i)~$\mu=1$, $b_0=1$, (ii)~$\mu=2$, $b_0=1$,
(iii)~$\mu=2$, $b_0=10$, (iv)~$\mu=5$, $b_0=10$. Other parameters 
are $a_0=1$, $k_b=1$, $T=4$. In each case $\kappa^2\sigma^2=
0.45\,\kappa^2\sigma^2_{\mathrm{max}}$. Note a~changing slope of 
the spectrum (iv).}
\label{fig-fcor}
\end{figure}

The Fourier transform of the autocorrelation function (\ref{asymptotycznie})
is by the Wiener--Khinchin theorem related to the power spectrum of the
process $\{a_n\}$, $n\gg 1$, averaged over realizations of the noise:

\begin{eqnarray}\label{power_n}
P(\omega) &=& v_\infty^2 \left|\sum\limits_{m=0}^\infty \alpha^m
e^{im\omega T} \right| 
\nonumber\\
&=& \sqrt{2}v_\infty^2 
\sqrt{\frac{1-\alpha\cos\omega T}{1 - 2\alpha\cos\omega T + \alpha^2}}\,,
\end{eqnarray}

\noindent where $\alpha = (1\nobreak+\nobreak{}b_\infty k_bT)^{-\mu}
\exp\left(b_0\kappa^2\sigma^2/(2k_b)\right)$. For all noise levels such 
that $v_\infty^2$ is finite the series in (\ref{power_n}) is convergent.
Because (\ref{power_n}) corresponds to the power spectrum of a discrete
time series ``sampled'' with a time step of $T$, there is a finite
Nyquist frequency $\pi/T$. If we want to probe higher frequencies, 
we need to examine correlations in the process $a(t)$ \textit{between} 
the influxes of the reagents.  It is sufficient to calculate $C(nT,\tau)$ 
with $\tau = mT + \tau^\star$, where $m$ is natural and 
$0\le\tau^\star<T$. For $n\gg1$, $C(nT,\tau) \simeq C(\tau)$ with

\begin{equation}\label{korelacja3}
C(\tau)=  
\frac{C_m}{(1 + b_\infty k_b\tau^\star)^\mu}
\exp\left({\textstyle\frac{1}{2}}\kappa^2\sigma^2
\frac{b_\infty^2\tau^\star}{1 + b_\infty k_b\tau^\star}\right).
\end{equation}

\begin{figure}
\includegraphics[scale=0.8]{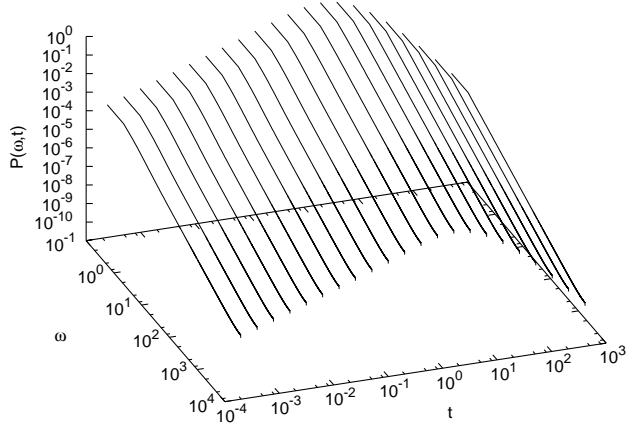}
\caption{Time-dependent spectrum of the system not sustained from
outside. For all times $t$, the power spectrum displays a clear power-law
behavior for large $\omega$. Parameters are $\mu=1$, $b_0=1$, $a_0=1$, 
$k_b=1$, $\kappa^2\sigma^2=1.98$. Units of $\omega$, $t$ are
reciprocal, but otherwise arbitrary.}
\label{fig-spcr}
\end{figure}

The correlation function (\ref{korelacja3}) corresponds to the power
spectrum of $a(t)$ averaged over realizations of the noise.
Typical spectra are plotted in Fig.~\ref{fig-fcor}. For low
frequencies, $\omega T<\pi$, which correspond to large $\tau$, the 
spectra are dominated by the behavior similar to (\ref{power_n}).
For very large frequencies (very small $\tau$), all curves plotted display
a clear $1/f$ (or $1/\omega$ in our notation) decay. This effect is
generic: If we expand (\ref{korelacja3}) in powers of $\tau^\star$, we get

\begin{equation}\label{expanded}
C(\tau)\simeq C_m\left[
1-b_\infty(\mu k_b- {\textstyle\frac{1}{2}}\kappa^2\sigma^2b_\infty)
\tau^\star\right] + O\left((\tau^\star)\right)\,.
\end{equation}

\noindent For very small $\tau^\star$, the correlation function approaches
a non--zero value, which is responsible for the $1/f$ decay at very large
frequencies. The second term in the expansion would lead to $1/f^2$ decay for
intermediate frequencies; however, as $\tau^\star$ cannot grow too large,
this effect can show up only if the coefficient is large enough
(cf.~spectrum (iv) on Fig.~\ref{fig-fcor}). If the time interval, $T$,
between the consecutive influxes is large, $\tau^\star$ can also be large and 
(\ref{korelacja3}) simplifies to

\begin{equation}\label{korelacja3a}
C(\tau)\simeq\frac{C_m\exp\left(\kappa^2\sigma^2b_\infty/2k_b\right)}
{(1+ b_\infty k_b\tau^\star)^\mu} \sim 
C_m(1 + b_\infty k_b\tau^\star)^{-\mu}\,.
\end{equation}

\noindent The coefficient in the above expression can be made
sufficiently large by choosing appropriate values of the parameters
of the system.

The spectra of 
Fig.~\ref{fig-fcor} become flat for very high frequencies, but this is 
a numerical effect resulting from the roundoff errors acting like
a white noise.

Let us compare the above results with the situation in which the reactions 
(\ref{A})--(\ref{B}) are not sustained from outside, $j_a=j_b=0$. Such
reactions can be experimentally realized in a closed reactor, as opposed
to the flow reactor considered above, but formally this  corresponds to taking
$T\gg1$ and considering the first interval only. 
Under these assumptions we obtain

\begin{subequations}\label{shorttime}
\begin{eqnarray}
\left\langle a(t)\right\rangle &=&
{\cal Z}(t)e^{\Psi(t)},
\\
C(t,\tau) &=& 
{\cal Z}(t){\cal Z}(t+\tau)
\left(e^{4\Psi(t) + \Phi(t,\tau)} - e^{\Psi(t)+\Psi(t+\tau)}\right),
\nonumber\\
\label{shorttime-cor}
\end{eqnarray}
\end{subequations}

\noindent where

\begin{eqnarray*}
{\cal Z}(x) = \frac{a_0}{(1+b_0k_bx)^\mu},\quad
\Psi(x) = {\textstyle\frac{1}{2}}\kappa^2\sigma^2\frac{b_0^2x}{1+b_0k_bx},\\
\Phi(x,y) = {\textstyle\frac{1}{2}}\kappa^2\sigma^2
\frac{b_0^2y}{(1+b_0k_bx)(1+b_0k_b(x{+}y))}.
\end{eqnarray*}

\noindent
If $\tau=0$, the second of equations (\ref{shorttime}) gives the
variance of the process $a(t)$. We can see that for
small $t$ the variance grows linearly, then reaches a maximum, and then
decays; for long times this decay has a power tail $\sim t^{-2\mu}$
regardless of the noise level: for times large enough almost all 
realizations of the process
$a(t)$ vanish, and therefore a statistical difference between these 
realizations vanishes as well. This is clearly related to the presence
and behavior of $b(t)$. Recall that a decrease in $b(t)$ limits the 
rate at which the substance A decays. For long times, when $b(t)\simeq0$,
even wide fluctuations of the reaction rate $K$ have little effect on
the actual decay of A. If the
concentration of B were kept constant ($k_b=0$), the expectation value
of $a(t)$, its variance and correlations described by (\ref{shorttime})
would either exponentially go to zero or expenentially diverge, depending
on the noise level.

\begin{figure}
\includegraphics[scale=0.72]{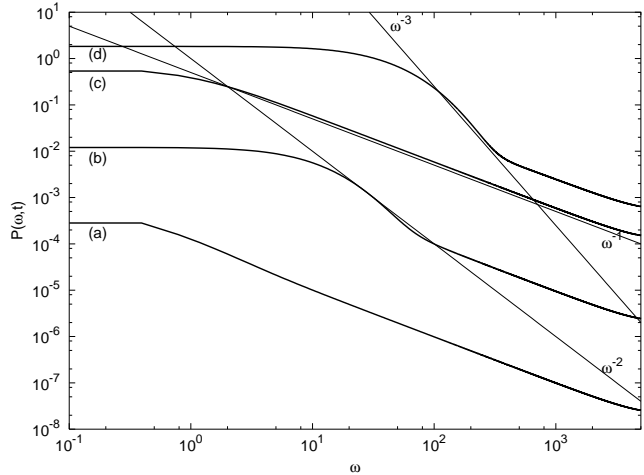}
\caption{Time--dependent spectra of the system not sustained from outside. 
Parameters are: (a)~$\mu=1$, $b_0=1$, $t=1/2048$, (b)~$\mu=5$, 
$b_0=10$, $t=1/2048$, (c)~$\mu=5$, $b_0=10$, $t=1/256$, (d)~$\mu=5$, 
$b_0=50$, $t=1/1024$. Other parameters, common for all the presented spectra, are: 
$a_0=1$, $k_b=1$, $\kappa^2\sigma^2=1.98$ (except for (d), where 
$\kappa^2\sigma^2=0.5$).
Parameters of the spectra (b), (c) are the same as those of
the spectrum (iv) of Fig.~\ref{fig-fcor}. Units of $\omega$, $t$ are
reciprocal, but otherwise arbitrary.}
\label{fig-slopes}
\end{figure}

Because in the not sustained case the process $a(t)$ does not reach
any stationary state, the correlation function (\ref{shorttime-cor})
depends on two arguments and the corresponding power spectrum, $P(\omega,t)$, 
is also
time--dependent. In (\ref{shorttime-cor}) we easily recognize the same
type of small $\tau$ behavior as in (\ref{korelacja3}) which leads to 
the $1/f$ tail for large frequencies. Indeed, this is what we
observe numerically (Fig.~\ref{fig-spcr}). Note that this power tail
does not change after averaging over the time, $t$:

\begin{equation}\label{timeaveraging}
\widehat P(\omega) = \frac{1}{\cal T}\int\limits_0^{\cal T} P(\omega, t) dt
\end{equation}

\noindent with ${\cal T}\gg0$ but finite, also displays this type of large
frequencies behavior. We may also expect a slope of $1/f^2$ type for 
intermediate frequencies and the parameters of the model large enough, but
as time, $t$,
increases, the frequency range corresponding to such a slope decreases, 
and for large (actually, not even \textit{very} large) values of $t$ the 
effect of the second order in $\tau$ becomes negligible 
(Fig.~\ref{fig-slopes}). For very large values of the parameters and small 
$t$, even the $1/f^3$ decay appears for intermediate frequencies, but this 
effect also disappears as time, $t$, grows. The presence of the $1/f^{1+n}$ 
slope is, therefore, at most a~transient effect, which does not survive time 
averaging~(\ref{timeaveraging}). In the limit $t\to\infty$, the correlation 
function (\ref{shorttime-cor}) becomes 

\begin{equation}\label{tinfty}
C(t\to\infty,\tau)\simeq \frac{\mathrm{const}}{(b_0k_bt)^{2\mu}}\,
(1+\tau/t)^{-\mu},
\end{equation}

\noindent which bears a formal similarity to
(\ref{korelacja3a}), but as the coefficient is very small, 
the $\tau$--dependence becomes practically unobservable. As we have said, 
this results from the fact that almost all realizations of the process $a(t)$ 
vanish in the limit $t\to\infty$ if the reactions are not sustained from outside.

If the concentration of B is kept constant and the concentration of
A is not incremented from outside,

\begin{equation}\label{justa}
C(t,\tau)=a_0\exp\left(-2b(k_a{-}b\kappa^2\sigma^2)t
-b(k_a{-}b\kappa^2\sigma^2/2)\tau\right).
\end{equation}

\noindent A similar expression can be found if the concentration of $\mathrm{A}$
is incremented from outside. There is no power--law dependence on $\tau$
in (\ref{justa}). It is now clear that the power-laws in (\ref{korelacja3a}) 
and (\ref{tinfty}) result from the fact that the stochastic process
$a(t)$ is driven by an ``external'', deterministic process $b(t)$.

In this Letter we have considered two cases: when the 
reactions are sustained from outside by periodic influxes of the reagents
and when they are not. In the former case, the effects of noise are partially
masked by the influxes, but since the noise effetcs in the latter cannot
be observed due to the eventual decay of the reagents (Eqns.~(\ref{shorttime})),
only the former admits a practical experimental realization. The mechanism 
responsible for the
emergence of the $1/f$ noise in the system under consideration is
following: kinetics of a proces driven by a multiplicative GWN
(the decay of A) is additionally driven by another, deterministic process
(the decay of B). 
To our knowledge, this is the first model in which a clear \textit{physical}
cause of the $1/f$ noise has been established. 
As no
power scalings or other special features have been assumed, even implicitly,  
our results suggest that this mechanism can have 
a more universal character. This idea will be pursued in a future research.

\end{document}